\documentclass[sigconf]{acmart}
\usepackage{xspace}

\newcommand\myname{KALM4REC\xspace}

\usepackage{amsmath}
\usepackage{multirow}
\DeclareMathOperator*{\argmax}{argmax}

\usepackage{xcolor}
\usepackage{subcaption}
\usepackage{listings}
\usepackage[normalem]{ulem}
\useunder{\uline}{\ul}{}

\usepackage[framemethod=TikZ]{mdframed}
\newcounter{theo}
[section]
\renewcommand{\thetheo}{}

\newcounter{lem}[section]
\renewcommand{\thelem}{\arabic{lem}}

\AtBeginDocument{%
  }
\copyrightyear{2025}
\acmYear{2025}
\setcopyright{acmlicensed}\acmConference[WWW Companion '25]{Companion Proceedings of the ACM Web Conference 2025}{April 28-May 2, 2025}{Sydney, NSW, Australia}
\acmBooktitle{Companion Proceedings of the ACM Web Conference 2025 (WWW Companion '25), April 28-May 2, 2025, Sydney, NSW, Australia}
\acmDOI{10.1145/3701716.3717855}
\acmISBN{979-8-4007-1331-6/2025/04}

\acmISBN{978-1-4503-XXXX-X/18/06}

\begin{document}

\title{Keyword-driven Retrieval-Augmented Large Language Models for Cold-start User Recommendations}

\author{Hai-Dang Kieu}
\affiliation{%
  \institution{VinUniversity}
  \city{HaNoi}
  \country{VietNam}
}
\email{dang.kh@vinuni.edu.vn}

\author{Minh-Duc Nguyen}
\affiliation{%
  \institution{VinUniversity}
  \city{HaNoi}
  \country{VietNam}
}
\email{duc.nm2@vinuni.edu.vn}

\author{Thanh-Son Nguyen}
\affiliation{%
  \institution{Institute of High Performance Computing,\\Agency for Science, Technology and Research}
  \country{Singapore}
}
\email{nguyen\_thanh\_son@ihpc.a-star.edu.sg}

\author{Dung D. Le}
\affiliation{%
  \institution{VinUniversity}
  \city{HaNoi}
  \country{VietNam}
}
\email{dung.ld@vinuni.edu.vn}

\renewcommand{\shortauthors}{Hai-Dang Kieu, Minh-Duc Nguyen, Thanh-Son Nguyen, and Dung D. Le}

\begin{abstract}
Recent advances in large language models (LLMs) have shown significant potential in enhancing recommender systems. However, addressing the cold start recommendation problem remains a considerable challenge. In this paper, we introduce a novel framework, namely KALM4Rec (\textbf{K}eyword-driven Retrieval-\textbf{A}ugmented Large \textbf{L}anguage \textbf{M}odels for Cold-start User \textbf{Rec}ommendations), designed to tackle this problem by using input keywords from users in a practical scenario of cold start user recommendations.
KALM4Rec operates in two main stages: candidates retrieval and LLM-based candidates re-ranking. In the first stage, keyword-driven retrieval models are used to identify potential candidates, addressing LLMs' limitations in processing extensive tokens and reducing the risk of generating misleading information. In the second stage, we employ LLMs with various prompting strategies, including zero-shot and few-shot techniques, to re-rank these candidates by integrating multiple examples directly into the LLM prompts.
Our extensive evaluation on two benchmarking datasets demonstrates that KALM4Rec excels in improving recommendation quality and also highlights its potential for widespread applications. Our code is available at https://github.com/dangkh/Kalm4rec-www
\end{abstract}


\maketitle

\section{Introduction}

Recommender systems are essential in assisting users navigating the vast number of available choices in the digital world. However, a significant and ongoing challenge in this field is addressing the issue of cold-start users. These are users who are new to the platform and therefore have no interaction history. The lack of data makes it difficult for the system to generate accurate and personalized recommendations. Conventional Collaborative Filtering (CF) such as  \cite{he2020lightgcn, liang2018variational, liu2023diffusion} struggle to suggest relevant items effectively for these new users due to the lack of detailed preference information. While user-user content-based algorithms offer a solution by employing user features to find similar users and recommend positively interacted items \cite{anwar2022collaborative, zhao2022improving, zhou2023contrastive}, this approach raises privacy concerns. Meanwhile, Large Language Models (LLMs) are gaining attention for enhancing recommender systems by leveraging their advanced language and reasoning abilities to address user needs \cite{wang2023rethinking} effectively. 

{\bf Challenges.} Despite their excellent capacities, existing LLMs suffer from several limitations when applied to recommender systems, especially for cold-start scenarios \cite{fan2023recommender, dai2023uncovering, geng2022recommendation, he2023large, sanner2023large, wang2024large}. For one, lacking comprehensive knowledge in a specific domain may result in nonfactual outputs. Although fine-tuning could potentially reduce the provision of irrelevant recommendations, this solution is often impractical due to the significant resources required \cite{mialon2023augmented}. Furthermore, directly incorporating user/item information can be costly in terms of token usage and restricts input length due to limited context length.
\begin{figure*}[ht!]
  \centering
  \includegraphics[width=0.6\textwidth]{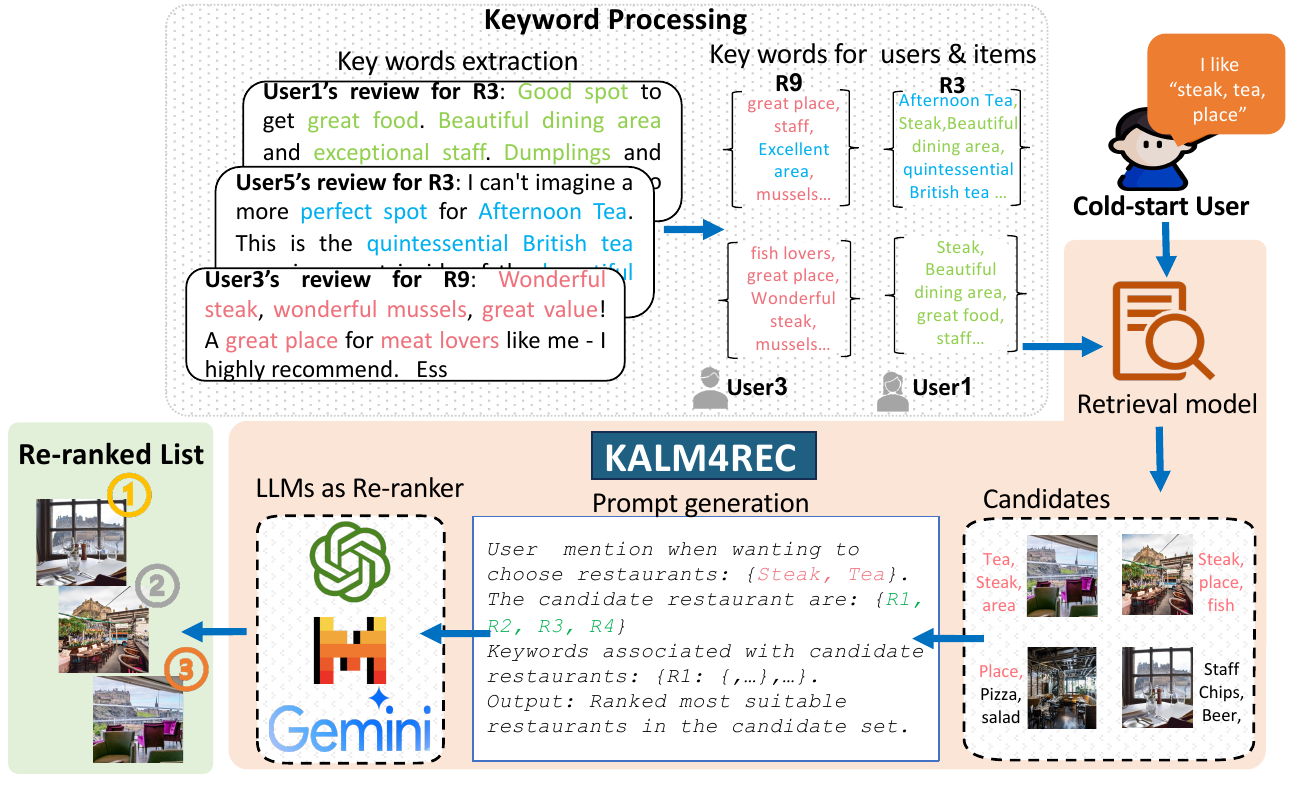}
  \caption{\myname begins by extracting noun phrases from user reviews to build profiles. Candidate restaurants are retrieved based on user's selected terms, then forwarded to the Prompt Generation, where user and restaurant details are incorporated. Finally, LLMs leverage their knowledge to capture user expectations and re-rank the candidates effectively.}
  \label{fig:pipeline}
\end{figure*}

{\bf Approach.} For recommender systems in the various domains, user reviews offer valuable information on their opinions about businesses. Modeling these personal reviews is beneficial to understand the preferences of users and capture the characteristics of restaurants \cite{vu2020multimodal}. For cold-start users who lack historical data, a practical strategy is to ask them to provide a few keywords that describe their preferences. This approach helps to create an initial user profile with minimal effort and without compromising their privacy. Furthermore, using keywords instead of entire reviews to prompt LLMs can enhance the efficiency and accuracy of recommendations while minimizing token usage.

To this end, we present a novel framework called \textbf{\myname} built upon two essential components centered around keywords: \textit{candidates retrieval} focusing on retrieving relevant items, and \textit{LLM-ranker } which leverages LLM to re-rank the retrieved candidates. 
User reviews reflect user preferences but often contain noise, potentially leading to inaccurate representations of user preferences. Using extracted keywords exclusively, we retain the contextual essence of the reviews while still describing the user profile well. Using keywords also allows us to optimize the prompts by reducing noise, thereby enhancing their effectiveness while still ensuring sufficient information of users and restaurants in LLMs, effectively addressing cold-start issues.
Essentially, \myname, grounded in sets of keywords, aims to narrow down the candidate set relevant to predefined cold-start preferences. The LLM utilizes its contextual understanding and reasoning capabilities to generate a ranked list of items for recommendations.

\section{Proposed Framework}

In this section, we present the details of our keyword-driven framework, \textbf{\myname} (Figure \ref{fig:pipeline}), designed to tackle the cold-start recommendation problem. \myname unfolds in two primary stages: (1) \textit{Candidates retrieval}, where meaningful noun phrases are extracted from collected reviews to form word sets representing users and targeted items, and subsequently utilized for queries from cold-start users to obtain potential candidates; and (2) \textit{LLM-based candidates re-ranking}, where an LLM is employed as a ranker to re-rank the retrieved candidates.
\subsection{Problem Formulation}
Given the set of users $\mathcal{U}$ and the set of targeted items (items for short) $\mathcal{R}$ and $i_{u,r}$ represents a review if user $u \in \mathcal{U}$ has a review for an item $r \in \mathcal{R}$. The task involves recommending relevant items to a new user $u_c$ (cold-start user). 
The new user can opt to declare their preference via a set of pre-defined keywords $k_{u_c}$. Simultaneously, user reviews also contain ``keywords'' with $k_{u,r}$ is a set of keywords extracted from $i_{u,r}$. We denote $k_{u}$, $k_{r}$ as a set of keywords extracted from all user $u$'s reviews, and  all reviews written for item $r$ while $\mathcal{K}$ represented for set of all extracted keywords in training data. The problem of keyword-driven cold-start user recommendation is defined as: \emph{Given a set of keywords $k_{u_c}$ from a cold-start (new) user, return a ranked list of relevant items $R_{u_c}$ where $R_{u_c} \subset \mathcal{R}$}. Next, we will present our \myname with the details. 

\subsection{Candidates Retrieval}
We use SpaCy\footnote{https://spacy.io/} to extract keywords from reviews by retaining consecutive words with specific part-of-speech tags: `ADJ', `NOUN', `PROPN', and `VERB'. These keywords capture diverse aspects mentioned in reviews. Throughout our study, both keywords and reviews are vectorized using sBERT\footnote{https://huggingface.co/sentence-transformers/all-MiniLM-L6-v2}. 

To retrieve candidates, we propose Message Passing on Graph (MPG), a heterogeneous graph with nodes representing keywords and items, and two types of edges. Additionally, we introduce a scheme for estimating the connection scores between keywords and items. Drawing inspiration from LightGCN \cite{he2020lightgcn}, we construct a graph using training data, with edges established for \textit{(user - keyword)} and \textit{(keyword - item)} interactions when a user uses a keyword to review and a item's review contains that keyword. 
Information is passed through the edges in the graph, and node information could be generated as follows:
\begin{equation}
q_r = AGG(q_w, w \in k_{r})
\end{equation}
where $AGG$ is a function that aggregates information from neighboring nodes;  $q_r, q_w$ denoted for node information of item $r$ and it's connected keywords $w$.

However, due to the challenge posed by a large number of nodes, reaching millions, we adopt an unsupervised learning approach. First, we adopt an aggregator $(AGG)$ based on a simple weighted sum (similar to the work from LightGCN \cite{he2020lightgcn}), but without trainable parameters to determine the weights. For an item node, information is generated by summing up the information of the adjacency keyword node as follows:
\begin{equation}
    q_r =\sum_{ w \in k_r} e^{w} * a^w_{r}  
\end{equation}
where $e^{w}$ denotes feature of keyword nodes $w$; $a^w_{r}$ denote the connection weight between node $w$ and node $r$. Specifically, $a^w_{r}$ can be represented by a value indicating the importance of a keyword to the item. We introduce a scheme named \textbf{TF-IRF} to measure the importance score of a keyword to an item, which is similar to the TF-IDF idea as below:
\begin{equation}
   a_r^w = tf_r^w * irf^w 
   \label{tfirf}
\end{equation}
with $tf_r^w = \frac{f_r^w}{q_w}$, where $f_r^w$ represents the number of times the term $w$ appears in the set of keywords for item $r$, and $q_w$ indicates the total number of times the term $w$ is used in the entire training data; $irf^w = \log \left(\frac{| \mathcal{R} | }{f^w}\right)$, where $f^w$ denotes the number of items containing the term $w$.
Then scores of all items $\mathcal{S}$ is estimated as:
\begin{equation}
    \mathcal{S} = \mathcal{M} \times \mathcal{A}
\end{equation}
where $\mathcal{S} \in \mathbb{R}^{1 \times |\mathcal{R}|},
    \mathcal{M} \in \mathbb{R}^{1 \times  |\mathcal{K}|} $ denotes the occurence matrix represents the connection between cold-start user and their selected keywords, matrix $\mathcal{A} \in \mathbb{R}^{|\mathcal{K}| \times |\mathcal{R}|} $ contains all score assigned to the edge established by a keyword $w$ and an item $r$.
Then, the top-k possible items are provided by $\mathcal{R}_{u_c} = \argmax_{k=20}(\mathcal{S})$. Notably, no training is needed for this approach, only graph construction. For cold-start users selecting keywords absent in the training data, we replace these with semantically similar alternatives. This is done by vectorizing keywords with a pretrained BERT model and using nearest-neighbor search to find the closest match \footnote{https://scikit-learn.org/stable/modules/neighbors.html}.

\begin{figure}[ht!]
    \centering    \includegraphics[width=1.0\linewidth]{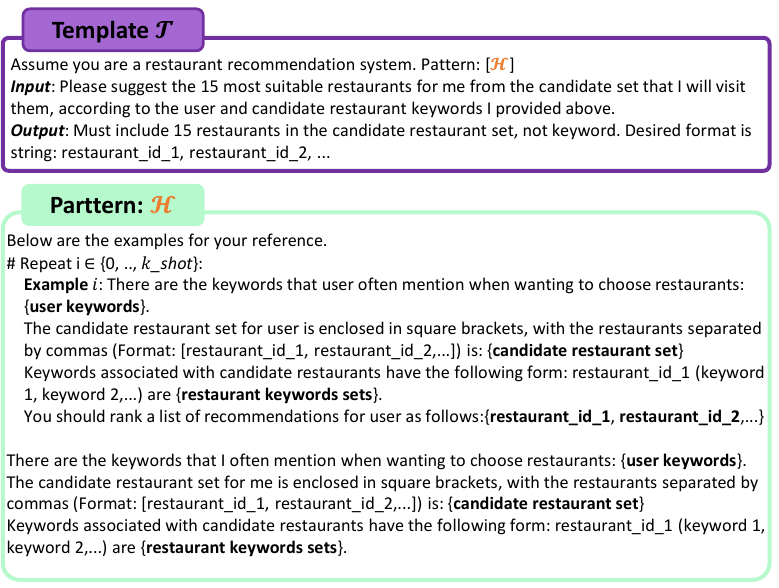}
    \label{fig:test1}
    \vspace{-0.1in}
    \caption{Our prompt template for re-ranking using keyword.}
\end{figure}
\noindent

\subsection{LLM-based Candidates Re-ranking}

We utilize LLMs to re-rank candidates for each user, followed by natural language instructions. Additionally, leveraging their reasoning and generation abilities, we incorporate information about user and item candidates into the instructions to make LLMs aware of user preferences. Like the retrieval model, user and item information is represented by keywords. We also include sentences in \textbf{Template $\mathcal{T}$} to trigger the recommender abilities of LLMs \cite{hou2024large} and to describe the task instructions to the model. Besides, we propose a general item recommendation prompt [\textcolor{orange}{$\mathcal{H}$}] pattern using keywords, consisting of (1) user keywords; (2) candidate sets; and (3) item keyword sets.
To better represent user interests and item characteristics, keywords are ranked by their \textbf{TF-IRF} scores. These keywords help LLMs capture nuanced user preferences and item attributes, enabling high-quality re-ranking. Like human reasoning, LLMs benefit from examples to interpret intentions and criteria more effectively. Therefore, our prompts are designed to utilize various prompting strategies, including zero-shot and few-shot techniques, by incorporating examples (selected from training users) within \textbf{Template $\mathcal{T}$}(referred to as Example i). For each selected example user, candidates are chosen based on keyword overlap, and the recommendation list is then ranked according to actual ratings.
In this section, we assess the candidate retrieval and the LLMs' capabilities in re-ranking candidates of \myname for cold-start user recommendations on a real-world dataset. We also investigate the impact of various factors on the quality of the LLMs' re-ranking.

\begin{figure*}[ht!]
\begin{subfigure}{0.19\textwidth}
        \centering
        \includegraphics[width=\textwidth]{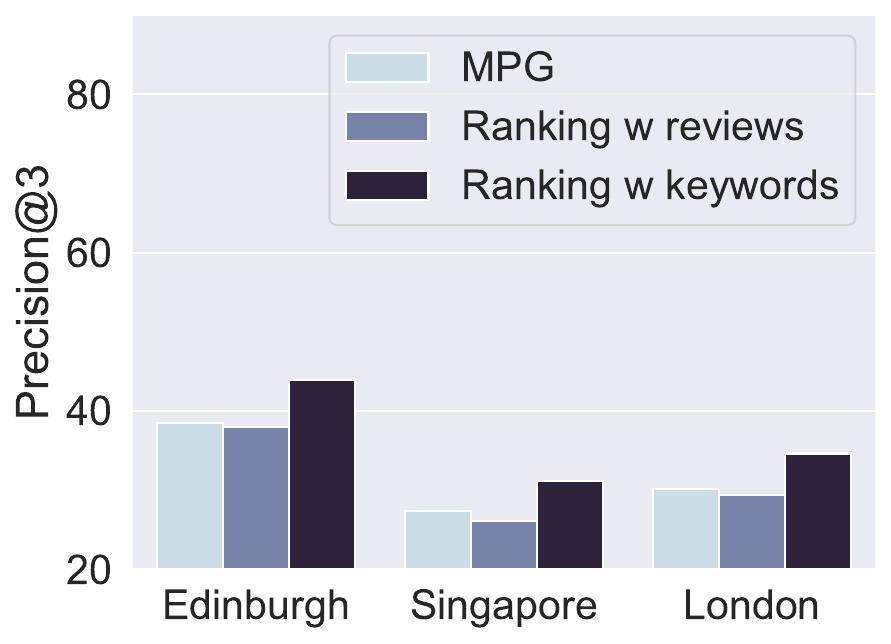}
        \caption{Review's performance}
    \end{subfigure}
\begin{subfigure}{0.19\textwidth}
        \centering
        \includegraphics[width=\textwidth]{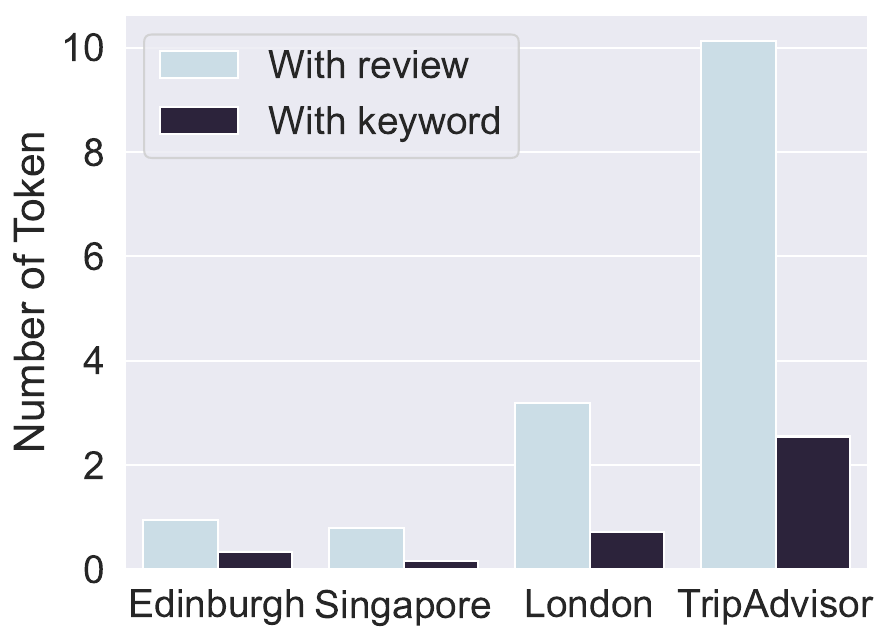}
        \caption{Number of token}
        \label{ablation_token}
    \end{subfigure}
\begin{subfigure}{0.19\textwidth}
        \centering
        \includegraphics[width=\textwidth]{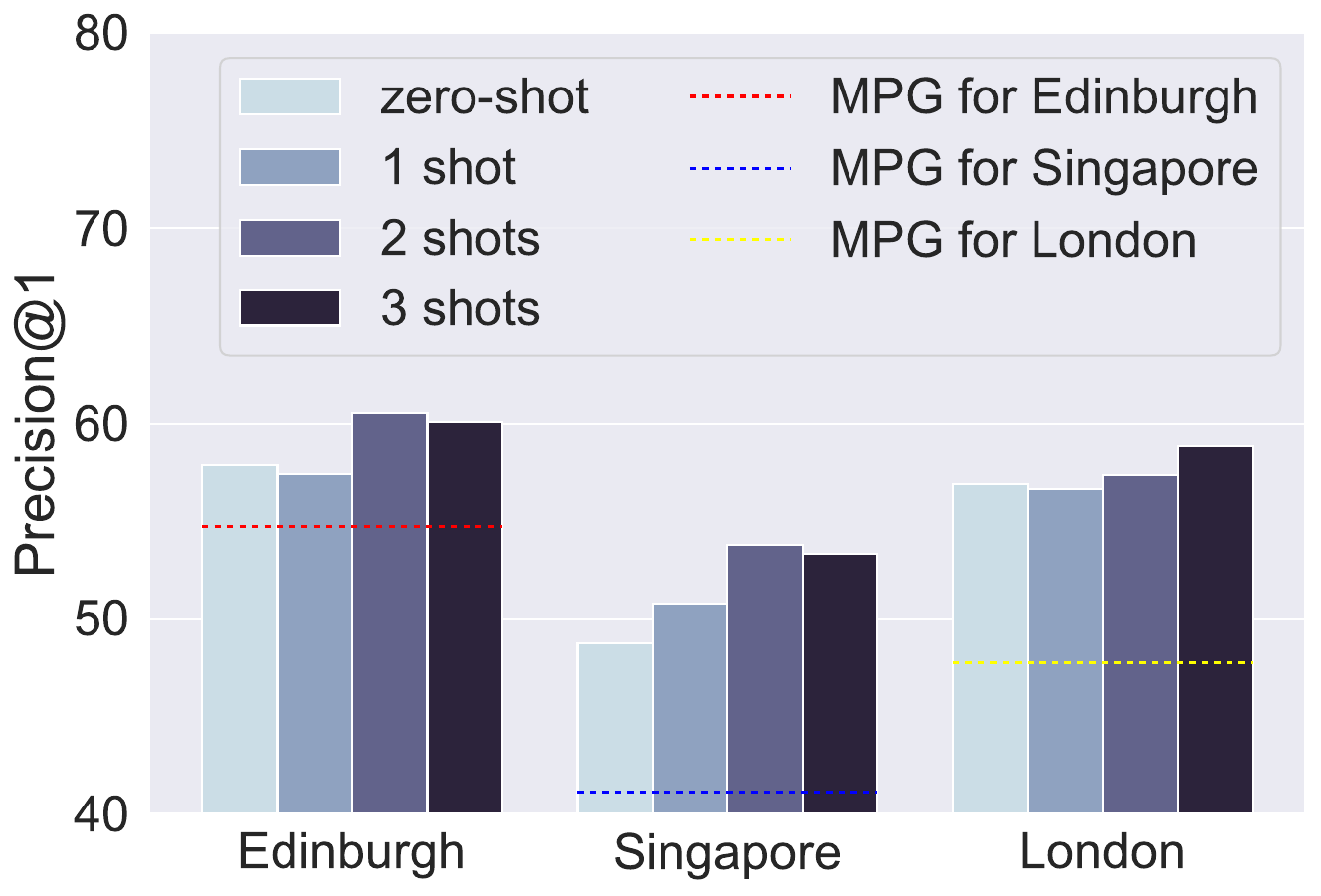}
        \caption{Fewshot performance}
        \label{grouped_shot}
    \end{subfigure}
\begin{subfigure}{0.19\textwidth}
        \centering
        \includegraphics[width=\textwidth]{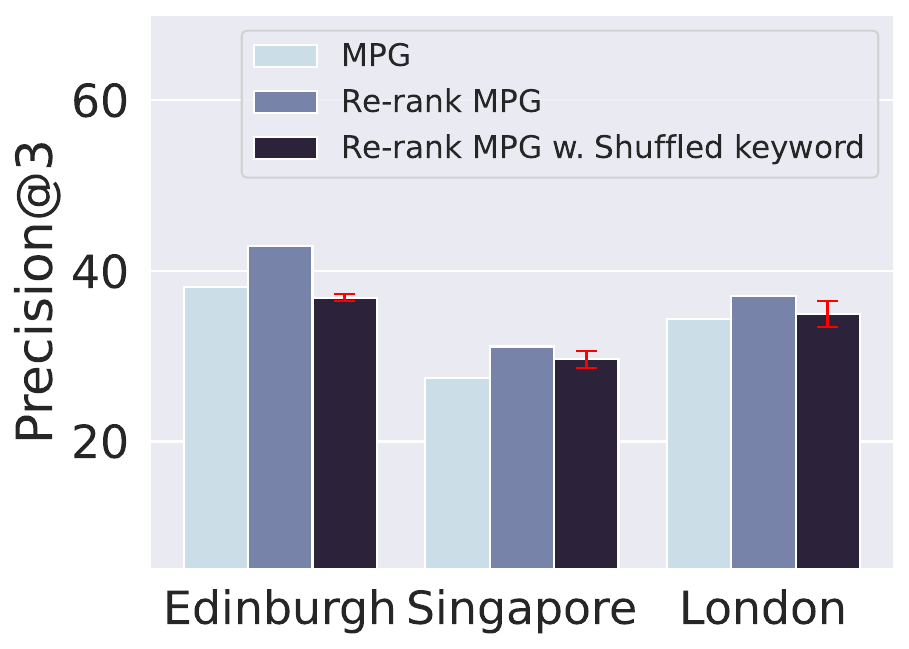}
        \caption{Keyword Order}
        \label{Keyword_Order}
    \end{subfigure}
    \begin{subfigure}{0.19\textwidth}
        \centering
        \includegraphics[width=\textwidth]{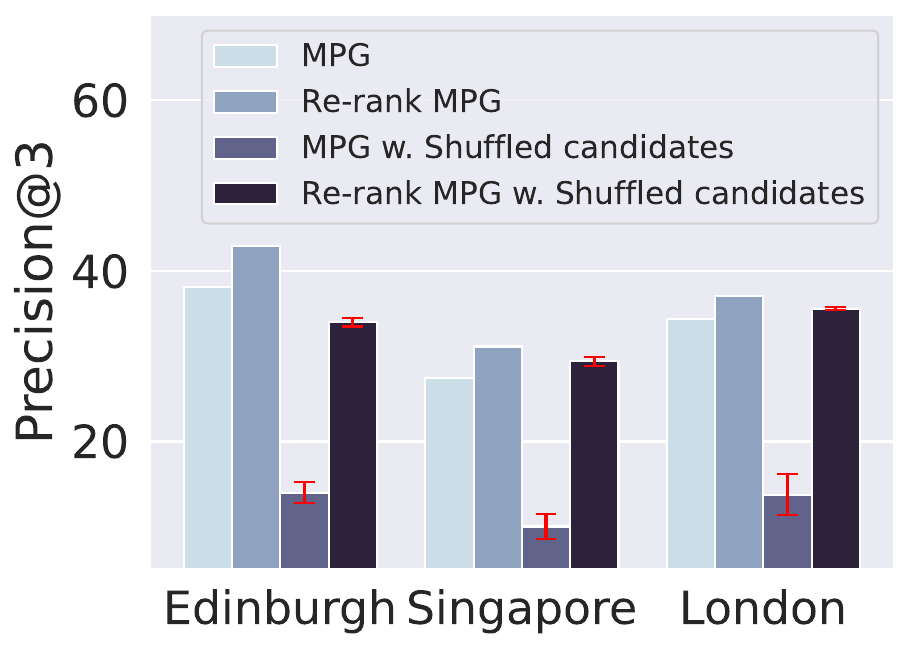}
        \caption{Candidate Order}
        \label{candidates_order}        
    \end{subfigure}
\label{fig:four_images}
\caption{Performance of \myname across various aspects using Gemini Pro }
\end{figure*}

\subsection{Experimental Settings}
\textbf{Dataset.} 
The experiments utilize two datasets: the Yelp.com dataset \cite{vu2020multimodal} and the TripAdvisor dataset \cite{li2013identifying}. The Yelp.com dataset features over 67,000 restaurant reviews across three English-speaking cities. The TripAdvisor dataset consists of 878,561 reviews from 4333 hotels crawled from TripAdvisor.com across 17 cities. Since we deal with the problem of cold-start user recommendation, we split the dataset so that users in the test set do not have any reviews in the train set. Dataset statistics are provided in Table \ref{tab2}\footnote{"Items" refer to restaurants in Yelp and hotels in TripAdvisor, respectively.}.

\begin{table}[ht!]
\caption{Dataset Statistics}\label{tab2}
\resizebox{1.0\columnwidth}{!}{%
{\setlength\doublerulesep{0.7pt}
\begin{tabular}{llrrrr}  
\toprule[1pt]\midrule[0.3pt]
\multirow{2}{*}{Dataset} & \multirow{2}{*}{City} & \multirow{2}{*}{\# Items} & \multicolumn{1}{c}{\multirow{2}{*}{\# Reviews}} & \multicolumn{2}{c}{\# Users}                         \\ \cline{5-6} 
& &                                 & \multicolumn{1}{c}{}                            & \multicolumn{1}{c}{Train} & \multicolumn{1}{c}{Test} \\
\midrule
\multirow{3}{*}{Yelp}& Edinburgh  & 938 & 12,753 &  1,187 &  223\\
& Singapore  & 983 & 12,878 &  1,046 &  197\\
& London     & 986 & 42,061 &  4,396 &  825\\
\hline
\multirow{1}{*}{TripAdvisors}&   & 4333 & 878,561 &  9,579 &  1,900\\

\midrule[0.3pt] \bottomrule[1pt]
\end{tabular}
}%
}
\end{table}

\noindent
\textbf{Evaluation and implementation details.}
We evaluate our model using three metrics: recall (R@K), precision (P@K), and Normalized Discounted Cumulative Gain (N@K). For the retrieval task, we focus on R@20 and P@20. For the re-ranking task, we choose various values of K, including 1, 3 are employed similar to recent works \cite{he2023large, hou2024large}.
Training, testing, and inference are conducted on Colab Pro with L4 GPU, batch size of 256 and AdamW optimizer with a learning rate of $1e^{-3}$. Experiments involving LLMs are conducted on Gemini Pro 1.5, GPT-3.5-Turbo, Mistral 8B, and LLama 3-8B. 

\begin{table}[ht!]
    \centering
    \caption{Performance of retrieval methods at P@20 (R@20).}
    \label{tab:retrieval}
\resizebox{1.0\columnwidth}{!}{%
{\setlength\doublerulesep{0.7pt}
\begin{tabular}{l|c|c|c|c|c}
\toprule[1pt]\midrule[0.3pt]
\multicolumn{1}{c|}{Dataset} & \multicolumn{1}{c|}{Jaccard} & \multicolumn{1}{c|}{MF} & \multicolumn{1}{c|}{MVAE} & \multicolumn{1}{c|}{CLCRec} & \multicolumn{1}{c}{MPG}      \\
\hline
Yelp-Edinburgh              & 4.32 (8.00)                 & 5.22 (13.66)           & 6.18 (17.5)              & 8.72 (27.59)                 & {  \textbf{14.42 (40.44)}} \\
Yelp-Singapore              & 3.47 (9.50)                 & 3.57 (11.51)           & 3.95 (13.48)             & 5.63 (20.87)                 & {  \textbf{10.38 (34.86)}} \\
Yelp-London                 & 2.58 (6.37)                 & 4.78 (15.49)           & 5.12 (16.15)             & 7.8 (26.74)                  & {  \textbf{12.73 (41.44)}} \\
TripAdvisor                 & 1.01 (2.65)                 & 7.37 (21.10)           & 5.10 (15.55)             & 7.22 (21.67)                 & {  \textbf{14.54 (43.74)}} \\ \hline
\end{tabular}
}%
}
\end{table}


\noindent
\textbf{Retrieval Baselines.}
We compared our MPG with retrieval methods: \textit{Jaccard similarity}, \textit{Matrix Factorization (MF)}, \textit{MVAE}, and \textit{CLCRec}. Jaccard similarity selects candidates based on overlapping keywords between users and items. For MF and MVAE, we identify similar users via Jaccard similarity and estimate item scores using their average ratings. For CLCRec \cite{Chen22}, we follow its proposed approach, using keywords as item content..

\noindent
\subsection{Experimental Results}
\textbf{Candidates Retrieval.}
The goal of candidates retrieval is not only to achieve a high ranking (precision) but also to retrieve as many correct candidates as possible (recall). This ensures that the re-ranking module has a comprehensive set of candidates to work with in its input. 
We evaluate the retrieval methods based on P@20 and R@20, which measure the precision and recall of the top 20 candidates, respectively. The results are presented in Table \ref{tab:retrieval},  our best retrieval model, MPG, compared to conventional methods. CLCRec performs better than all others conventional method as it utilized keywords to compute user and item representations. The best performance is achieved using MPG, which harnesses the graph structure between keywords and items. MPG demonstrates consistent improvement across three cities. 


\begin{table}[h!]
    \centering
    \caption{Performance of \myname using different LLMs for re-ranking with MPG as the retrieval method.}
\resizebox{0.9\columnwidth}{!}{%
{\setlength\doublerulesep{0.7pt}
\begin{tabular}{l|c|ccccc}
\toprule[1pt]\midrule[0.3pt]
City Name                  & Model                          & \multicolumn{1}{c}{P@1} & \multicolumn{1}{c}{R@1} & \multicolumn{1}{c}{P@3} & \multicolumn{1}{c}{R@3} & \multicolumn{1}{c}{N@3} \\ \hline
\multirow{5}{*}{Yelp-Edinburgh} & MPG                            & 54.71                   & 10.91                   & 38.12                   & 20.55                   & 58.49                   \\ \cline{2-7}
                           & Gemini                         & \textbf{61.69}          & \underline{11.54}             & \textbf{42.95}          & \textbf{21.36}          & \textbf{62.19}          \\
                           & GPT-3.5                        & 56.05                   & 11.21                   & \underline{38.86}             & \underline{21.26}             & 58.72                   \\
                           & Mistral 8B                             & \underline{58.30}             & \textbf{11.85}          & 37.82                   & 20.87                   & \underline{60.16}             \\
                           & \multicolumn{1}{l|}{Llama3 8B} & 56.50                   & 11.27                   & 36.62                   & 18.91                   & 56.42                   \\ \hline
\multirow{5}{*}{Yelp-Singapore} & MPG                            & 41.12                   & 8.19                    & 27.41                   & 15.55                   & 48.11                   \\\cline{2-7}
                           & Gemini                         & \textbf{51.78}          & \textbf{9.77}           & \textbf{31.13}          & \textbf{16.38}          & \underline{51.05}          \\
                           & GPT-3.5                        & 48.73                   & 8.85                    & \underline{30.29}             & 16.22                   & \textbf{51.86}             \\
                           & Mistral 8B                             & \underline{49.75}                   & \underline{9.67}           & \underline{30.29}             & \underline{16.27}             & 50.02                   \\
                           & \multicolumn{1}{l|}{Llama3 8B} & 45.18                   & 8.71                    & 29.10                   & 14.99                   & 49.27                   \\ \hline
\multirow{5}{*}{Yelp-London}    & MPG                            & 47.76                   & 9.60                    & 34.34                   & 19.47                   & 52.46                   \\\cline{2-7}
                           & Gemini                         & \textbf{58.06}            & \textbf{11.77}           & \textbf{37.09}            & \textbf{20.67}            & \textbf{56.85}         \\
                           & GPT-3.5                        & 52.97                   & \underline{10.81}             & 34.00                   & \underline{19.38}             & 53.96                   \\
                           & Mistral 8B                             & \underline{53.82}             & 10.79                   & 33.74                   & 18.87                   & \underline{54.21}             \\
                           & \multicolumn{1}{l|}{Llama3 8B} & 51.64                   & 10.39                   & \underline{34.08}             & 18.83                   & 53.49             
\\ \hline
\multirow{2}{*}{TripAdvisor}    & MPG                            & {54.60} & {8.73} & {41.92} & {19.72} & {61.36}                  \\\cline{2-7}
                           & Gemini                         & \textbf{59.43}            & \textbf{9.54}           & \textbf{42.15}            & \textbf{19.87}            & \textbf{62.36}         \\                           \midrule[0.3pt] \bottomrule[1pt]
\end{tabular}
}%
}

\label{tab:rerank}
\vspace{-0.1in}
\end{table}

\noindent
\textbf{Re-ranking Capability.} This section evaluates the re-ranking capabilities of \myname under various conditions. First, we investigate whether LLMs can improve recommendations for cold-start users. Results show that combining retrieval with LLM-based re-ranking in our \myname framework consistently outperforms retrieval-only methods, with Gemini achieving the best precision and recall (Tables \ref{tab:rerank}). Next, we explore using keywords instead of full reviews due to context length limitations, finding that keywords not only boost performance but also reduce token costs, enhancing efficiency (Figure \ref{ablation_token}). We then examine how prompt design, specifically zero-shot and few-shot strategies, impacts effectiveness. Few-shot prompts with more examples (e.g., 3-shot) yield the best results as they help LLMs better capture user intent (Figure \ref{grouped_shot}). Finally, we assess potential biases in LLM ranking, showing that keyword and candidate order significantly affect outcomes. Ordered keywords and optimized candidate order yield better performance, especially when paired with retrieval models (Figure \ref{Keyword_Order},\ref{candidates_order}). 
\section{Discussion and Conclusion}
In this work, we investigate the idea of augmenting LLMs for cold-start user recommendations with keywords extracted from user reviews. 
For retrieving potential candidates, we present MPG, keyword-based methods, which outperform conventional approaches. We then employ LLMs to re-rank the obtained candidates using designed prompting strategies that incorporate keywords to represent users and items. Comprehensive experiments indicate that \myname is capable of handling cold-start user scenarios effectively. Additionally, our framework shows a potential of integrating different retrieval and language models to achieve promising performance under various factors in the future for multiple domains.


\section*{Acknowledgment}
This research was funded by VinUniversity Seed Grant under project code 400088.

\bibliographystyle{ACM-Reference-Format}
\bibliography{main}


\end{document}